\documentclass[aps,prb,preprint,showpacs,superscriptaddress,floatfix]{revtex4}

\usepackage{graphics}
\usepackage{amsmath}
\usepackage{amssymb}
\usepackage{calc}
\usepackage{epsfig}
\usepackage{color}

\newcommand{\degre}{\ensuremath{^\circ}}

\begin{document}

\title{ Geometry and Surface Plasmon energy }

\author{R. Vincent}
\email{remi.vincent@espci.fr}\affiliation{Institut Langevin, ESPCI ParisTech, CNRS, 10 rue Vauquelin,
75005 Paris, France}
\author{J. I. Juaristi}
\affiliation{Donostia International Physics Center DIPC,
Paseo Manuel de Lardizabal 4,
Donostia-San Sebasti\'an 20018, Spain}
\affiliation{Departamento de F\'{\i}sica de Materiales,
Facultad de Qu\'{\i}micas,
UPV/EHU, Apartado 1072, 20080 San Sebasti\'an, Spain}
\affiliation{Centro de Física de Materiales (CSIC-UPV/EHU) — Materials
Physics Center MPC, P. Manuel de Lardizabal 5,
20018 San Sebastián, Spain }
%
%
\author{P. Apell}
\affiliation{Donostia International Physics Center DIPC,
Paseo Manuel de Lardizabal 4,
Donostia-San Sebasti\'an 20018, Spain}
\affiliation{Chalmers University of Technology,
Department of Applied Physics, SE-41296 G\"oteborg, Sweden}

\date{\today}

%
%

\begin{abstract}
We derive a simple rule to determine surface plasmon
energies, based on the geometrical properties of the surface of the metal. We
apply this concept to obtain the surface plasmon energies in wedges,
corners and conical tips.
The results presented here provide simple and straightforward rules
to design the energy of surface plasmons in severals situations of
experimental interest such as in plasmon wave guiding and in
tip-enhanced spectroscopies.


 \end{abstract}

\maketitle

\section{Introduction}

Plasmonics has become an important branch of nanooptics that allows
for the manipulation of light in the nanoscale thanks to localized
surface collective modes so called surface plasmons. Surface-plasmon propagating in plasmonic
circuits \cite{Devaux06} may act as interfaces in electrooptical
devices, and localized surface plasmons play a key role as the
mechanism of field-enhancement in field-enhanced spectroscopies
assisted by metallic nanoantennas \cite{Segerink08}. The capacity to
tune the optical response in metallic systems results of crucial
interest in many processes of
physics \cite{Prodan03}, chemistry \cite{Ward07}, biology 
\cite{Szmacinski95,Aslan05} and medicine \cite{Lal08} where
metallic systems act as active hosts responsible for the existence
and efficiency of the processes.

The energy  of surface plasmon is known to depend on the metallic
material, the geometry of the system, and the environment. Many
different geometries have been studied from the simple metallic
sphere or spheroids \cite{Mie08} treated analytically in the
beginning of last century to complicated coupled systems tackled
with use of sophisticated numerical methods to solve Maxwell's
Equations \cite{Abajo97,Farjadpour06,Yurkin07}.
 Here we derive
a simple rule to determine surface plasmon energies in direct
connection with the metallic shape.
 Laplace equation is
expressed in terms of a surface charge density at the boundaries of
the metal surface, and solved using a standard method of integral
equation. This method allows for straightforward connection between
 the surface plasmon energies and the solid angle at the metal-vacuum interface.

The simplicity of the concept allows for direct estimation of
plasmon energies in common situations of interest such as in a wedge
waveguide with plasmons running along a metallic wedge, or at the
tip of a metallic cone where surface plasmons are
localised at the tip apex. This piece of information permits to estimate and design
the energy of surface plasmon that can be later fine-tuned to
engineer and control their functionality.
\begin{figure}
\centerline{\resizebox{09.cm}{!}{\includegraphics{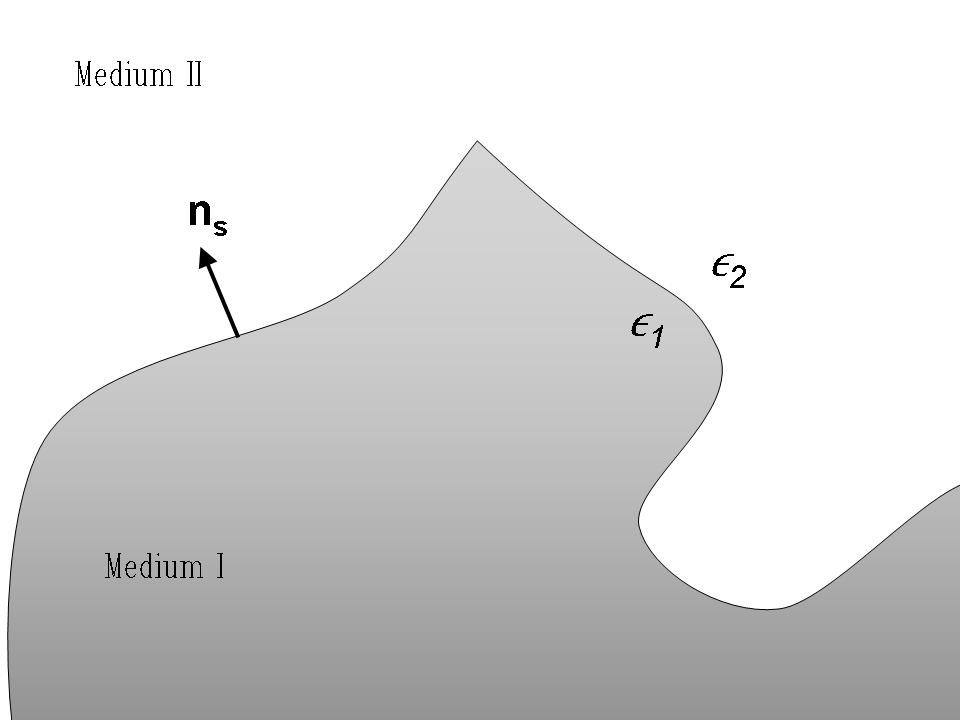}}}
\caption{\label{fig_shape}Interface between two regions with
different dielectric constants $\epsilon_1$ and $\epsilon_2$ bounded
by a surface S. }
\end{figure}

\section{Extended Theory}

Here we investigate the electromagnetic modes localized at the
surface $S$ separating two different media I-II, characterized by two
dielectric constants $\epsilon_1$ and $\epsilon_2$ respectively.
The normal vector ${\bf
n_s}$ at the interface position ${\bf s}$ has been chosen arbitrarily
to point towards medium II, see FIG. 1. In this way,
we shall call "internal" the variable defined in the part of space corresponding to medium I, and
"external" the variable defined in the part of space corresponding to medium II. For instance, $\epsilon_1$ will be denoted in the following by $\epsilon_{int}$ and $\epsilon_2$ by $\epsilon_{ext}$.

We want to obtain the surface plasmon modes of this
system. These modes can be
described by the Laplace equation for the quasi-static potential  $u({\bf r})$ of the self field,
 everywhere
except at the surface $S$ where we have the following equation
\begin{equation}
\nabla^2 u({\bf r})  = -\sigma({\bf r}), \label{laplacesurf}
\end{equation}
with $\sigma({\bf r})$ the surface charge density, i.e. different from zero on the contour surface $S$. Let us note by $ {\bf x}$ the position vectors located on the surface, therefore $\sigma ({\bf x})$ is related to the normal
derivatives of the potentials internal $u_{int}$ and external $u_{ext}$
to the surface $S$ by the following relation \cite{Abajo02}
\begin{eqnarray}
\label{eqdensurf} \sigma({\bf x}) =\frac{\partial u_{int}}{\partial
n_{x}}- \frac{\partial u_{ext}}{\partial n_{x}}.
\end{eqnarray}
The continuity equation
 for the self-fields at the boundary gives the following condition between the normal derivatives of the potential at any point of the surface
$S$
\begin{eqnarray}
\epsilon_{int} \frac{\partial u_{int}}{\partial
n_{x}}=  \epsilon_{ext} \frac{\partial u_{ext}}{\partial n_{x}}.
\end{eqnarray}
From the previous equations, we can recast Eq. (1) as follows
\begin{eqnarray}
u({\bf x})=\int_S G({\bf x},{\bf x'})\sigma({\bf x}') dS({\bf x}')\nonumber\\
=\int_S G\left(\frac{\partial u_{int}}{\partial n} -\frac{\partial u_{ext} }{\partial
n}\right) dS({\bf x}')\nonumber\\
=(1-\epsilon_{int}/\epsilon_{ext})\int_S G\frac{\partial u_{int}}{\partial n} dS({\bf x}')
\end{eqnarray}
where $G({\bf x},{\bf x'})=1/4\pi |{\bf x}-{\bf x'}| $ is the Green function solution of the homogenous Laplace equation
\begin{eqnarray}
\nabla^2 G({\bf x},{\bf x'})=-\delta ({\bf x}-{\bf x'})
\end{eqnarray}
where $\delta$   is the Dirac delta function.

Now considering a point ${\bf x}={\bf x}_o$ on the boundary S with possible singularities, and applying eqs. (A5) and (A6) of the appendix, i.e. evaluating the integral in Eq. (4) in Cauchy Principal Value sense, we obtain the following integral equation\cite{vincent09}
\begin{eqnarray}
u({\bf x}_0)\left[ \frac{\Omega({\bf x}_0)}{4\pi}- \frac{\epsilon_{ext}}{\epsilon_{ext}-\epsilon_{int}}\right]\nonumber\\
=\int_S u({\bf x}')\left( -\frac{\partial G }{\partial n}({\bf x}_0,{\bf x'})\right) dS({\bf x}')
\end{eqnarray}
where $\Omega({\bf x}_0 )$   is the solid angle sustained by the surface at the point  ${\bf x}_0$.

Let us measure solid angles in units of $4\pi$($\hat{\Omega}\equiv\frac{\Omega}{4\pi}$) and define $\delta \hat{\Omega}= \hat{\Omega}-1/2$  as the deviation from a smooth surface. Notice that $ \delta \hat{\Omega}<0$ for a surface with a protrusion (convex) and  $ \delta \hat{\Omega}>0$ for a concave surface. We can now write
\begin{eqnarray}
 \int_S d\hat{\Omega}({\bf x}_0,{\bf x'}) u({\bf x'})= \lambda u({\bf x}_0)
\end{eqnarray}
where we have used the fact that $ \int_S \left( -\frac{\partial G }{\partial
n}\right) dS({\bf x}')=\frac{\Omega({\bf x}_0)}{4\pi}$, and consequently $d\hat{\Omega}({\bf x}_0,{\bf x}')$ is the normalized elementary solid angle sustained by $dS({\bf x}')$ viewed from ${\bf x}_0$. To the best of our knowledge, the surface integral equation allowing to obtain the surface modes of the corresponding system was never written as a solid angle integral equation Eq.(7). This equation is pointing out the purely geometrical nature of these modes. We note that equation (7) is an exact expression, that complements recent numerical approximations \cite{Mayergoyz05} for smooth surfaces, and puts them into a more general context that includes surfaces with singularities.

Finally we interpret equation (7) as an eigenvalue equation where
\begin{eqnarray}
\lambda\equiv[\lambda_0(\omega)+\delta\hat{\Omega}]			
\end{eqnarray}
and we have defined the frequency factor
\begin{eqnarray}
 \lambda_0(\omega)=\frac{\epsilon_{ext}+\epsilon_{int}}{2(\epsilon_{int}-\epsilon_{ext})}=1/2-(\omega/\omega_b)^2.
\end{eqnarray}
In the last line we have introduced the generic case of a Drude metal for $\epsilon_{int}$ with bulk plasma frequency $\omega_b$ and vacuum value for $\epsilon_{ext}$. Notice again that the right hand side of Equation (6) is entirely a geometric object. Furthermore, since the integral is evaluated in the Cauchy Principal Value sense its main singularity is already present in the factor $\delta\hat{\Omega}=\hat{\Omega}-1/2$.

Now let us consider the $\lambda=0$ eigenvalue

\begin{eqnarray}
\epsilon_{ext}(1-\hat{\Omega})+\epsilon_{int}\hat{\Omega}=0
\end{eqnarray}

In the Drude limit of Eq.(9), we obtain the following for the plasmon energy $\omega$
%
\begin{eqnarray}
(\omega/\omega_b)^2=\hat{\Omega}.
\end{eqnarray}

\section{Results and discussion}

The derivation of the previous section has shown the crucial connection between the shape and the spectral properties of the surface plasmon modes.
In this section,  for some selected geometries, we demonstrate the existence of the $\lambda=0$ eigenmode.
Afterward, we apply Eq. (10) corresponding to the $\lambda=0$ case for metallic-vacuum interface, using a Drude dielectric response for the metal, and obtain the energies of these modes.

\subsection{$\lambda=0$ mode for a semi-infinite metal}

The planar interface is a well known geometry for its analytical solutions of the Laplace equation.
For a semi-infinite metal, with ${\bf x}_0$ on the planar surface, any elementary solid angle of the surface, view from ${\bf x}_0$ vanishes ($d\hat{\Omega}({\bf x}_0,{\bf x}')=0$). Consequently any function defined only in the interface, will be solution of the integral equation (7), with eigenvalue $\lambda=0$. For a planar surface, as we are dealing with a smooth surface, we have $\hat{\Omega}=1/2$. Therefore applying Eq. (11), we obtain the well known infinitely degenerated surface plasmon mode of frequency $\omega_s=\omega_b/\sqrt{2}$.

\subsection{Modes localized in one dimensional surface singularities}

Now let us imagine a transformation, which generates from a smooth interface an interface with singularities, e.g., folding the previous planar interface along a straight line. In this way, we create a one dimensional straight singularity. Along the singularity, we can define a geometric solid angle. For example, let us consider a wedge of 90 degrees. In this case, the solid angle sustained by the surface at the singularity is $\pi$ ($\hat{\Omega}=1/4$), for the modes localized along the singularity, i.e.
defined on the singularity only, the surface integral in Eq. (7) reduces to a line
integral over the singularity. Any function defined along this straight singularity, only and anywhere else,  will be solution of the line integral equation with the eigenvalue $\lambda=0$. Therefore using a Drude approximation of the dielectric constant of the metal and using Eq. (11), we obtain a plasmon wedge frequency $\omega_s=\omega_b/2$.

This derivation can be extended to a general wedge solid angle. Let it be $\alpha$ the interior angle defined by the intersection of two semi-infinite planes (see Fig. 2). In this case the solid angle sustained by the surface at the singularity is $2\alpha$ ($ \hat{\Omega}=\alpha/2\pi$). Using Eq. (11), we obtain the plasmon wedge frequency $\omega_s=\omega_b\sqrt{\alpha/2\pi}$.
\begin{figure}[htp]
\centerline{\resizebox{7.5cm}{!}{\includegraphics{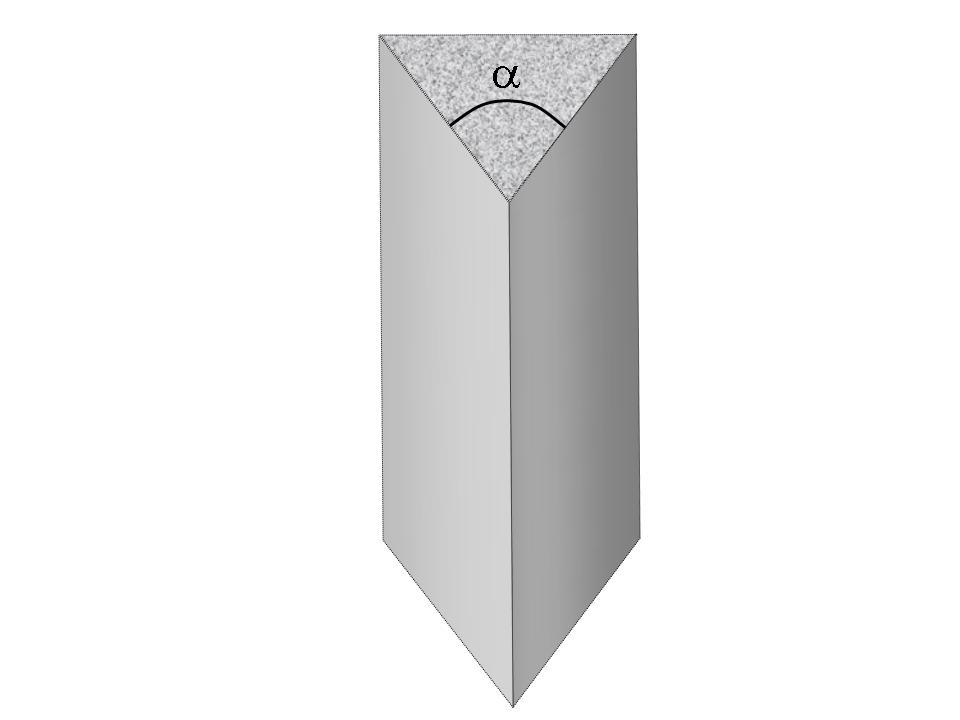}}}
\caption{Wedge geometry with an interior angle
$\alpha$.\label{fig_wedge}}
\end{figure}

\subsection{Modes localized at a sharp point}

Let us consider a point surface singularity, i.e.,
a metal vacuum interface with a sharp point.
In this case, in Eq. (7), the integral over the singularity vanishes, giving also for the mode localized at the
singularity, a $\lambda=0$ mode as the only possible solution.

For the conical geometry, where the full metallic cone
is defined by the
 apex angle $2\alpha$, illustrated in figure Fig. (3),
we calculate using Eq. (11), the frequency of the mode
localized at the apex. The geometric solid angle is
given by $\Omega_s=2\pi(1-\cos(\alpha))$, giving for the frequency of the mode, called here a dot plasmon $\omega_b$ mode:
\begin{eqnarray}
\omega_d^2=\frac{1-\cos\alpha}{2}\omega_b^2
\end{eqnarray}
It is interesting to notice that for small angle, this mode becomes linearly angle dependent, $\omega_d \sim \omega_b\alpha/2 $.
These results are relevant due to the important role of the tip geometry, for the
tip-enhanced spectroscopies, when the tip can can be modelized by a metallic cone.
\begin{figure}[htp]
\centerline{\resizebox{7.5cm}{!}{\includegraphics{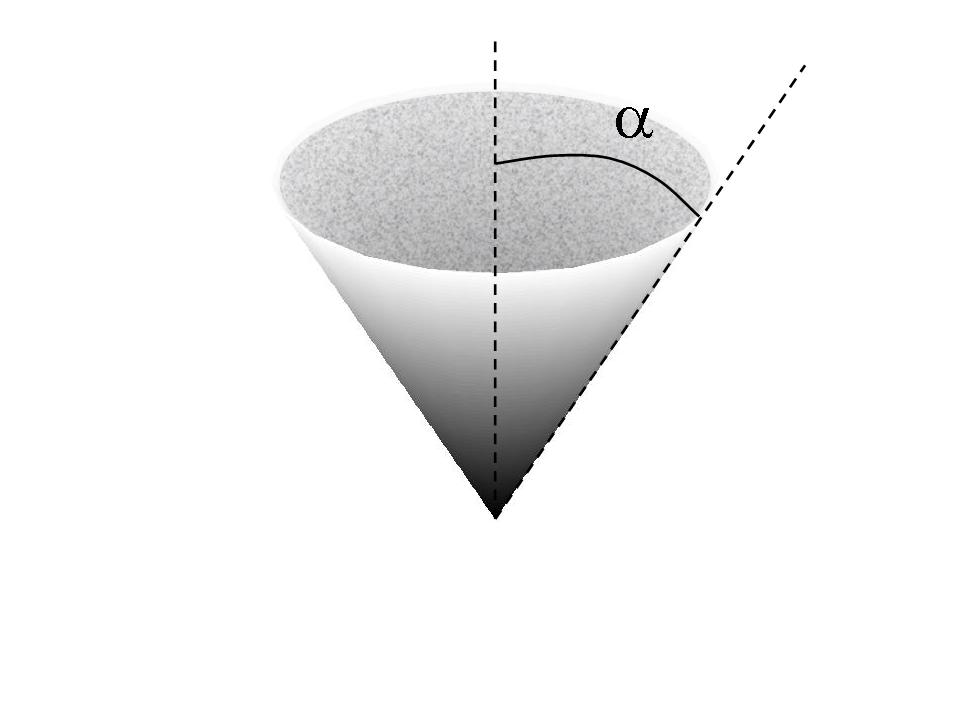}}}
\caption{Conical geometry with an apex angle $2\alpha$.\label{fig_cone}}
\end{figure}

These results are generalizable to modes
localized at corners. Let us consider the
corner at the intersection of three semi-infinite planes normal to
each other, where the metallic part is filling $1/8$ of the space.
We have at the intersection of the planes the solid angle of
$\Omega_s=\pi/2$. Using equation (11), we
obtain the frequency of the mode localized at the corner,
\begin{eqnarray}
\omega_d=\frac{1}{2\sqrt {2}}\omega_b.
\end{eqnarray}

\subsection{Complementarity rule}

One of the important properties of the plasmon modes demonstrated in smooth surfaces is the complementarity rule that relates the plasmon energies of a system and those of its complementary \cite{Apell96}.
Here, we are going to demonstrate that it can be generalized to surfaces with singularities.
The important point, as it is obvious from Eq. (7), is that the eigenvalues
of a system $\lambda(S)$ and the ones of its complementary
$\lambda(\overline{S})$ are the same: \cite{vincent09}
\begin{eqnarray}
\lambda(\overline{S}) =\lambda(S)\label{}.
\end{eqnarray}
The reason is that the two systems show the same surface, the only difference is that the values of the dielectric constant at both sides of the surface are interchanged. In other words, to treat the complementary system, we just need to
interchange the values of $\epsilon_{int}$ and $\epsilon_{ext}$ in Eq.
(10).
For a Drude metal, one finally obtains
\begin{eqnarray}
\omega_i^2(S)+\omega_i^2(\overline{S})=\omega_b^2.
\end{eqnarray}
This rule is obviously applicable to any surface
geometry, i.e., no matter it is a smooth surface or with a sharp
singularity where a solid angle can be defined.

\subsection{Extension to general dielectric materials}
If the metal is not a Drude metal and/or the second medium is not vacuum, one can still get the plasmon frequencies using the more general equation(10). In this case, one can get the plasmon frequencies using experimental data for the dielectric function of both systems\cite{Palik85}.

In table I, we summarize for the different cases explained above, the results for the plasmon frequencies obtained for Drude metals, and the conditions to be fulfilled by the ratio of the dielectric function at the plasmon frequencies when the more general (10) has to be used.


\begin{table}
\begin{tabular}{|l|c|c|c|c|c|c|}
  \hline
   && Semi-infinite & Wedge($90\degre$) & Wedge($\alpha$) & Corner($90\degre$) & Dot plasmon($\alpha$) \\
   &&  medium & &  &  & conical  \\
  \hline
   $(\omega/\omega_b)^2$ && 1/2 & 1/4 & $\alpha/2\pi$ & 1/8 & $(1-\cos \alpha)/2$\\
  \hline
  $\epsilon_{int}/\epsilon_{ext}$ && -1 & -3&$1-2\pi/\alpha$ & -7 & $(\cos \alpha+1)/(\cos \alpha-1)$ \\
  \hline

\end{tabular}
\caption{ Comparison  for different geometries of surface singularities squared of the plasmon energy divided by the bulk plasmon energy within the Drude model (see Eq. 11). Values of the ratio $\epsilon_{int}/\epsilon_{ext}$, between the interior and the exterior dielectric functions at the frequency of the surface plasmon for different geometries of surface singularities (see Eq. (10)).
}
\end{table}

\section{Summary}

Herein, we have derived a general framework (7-9) and simple geometrical rules Eq. (10-11) which determines
surface plasmon energies and plasmon modes, based on the shape of the metal. The
results presented here provide simple and straightforward rules to
design the energy of surface plasmons in several situations of
experimental interest such as in plasmon wave guiding and in
tip-enhanced spectroscopies.
 We have shown applications of the rule to obtain the surface plasmon
energies in wedges, corners and conical tips.

If the dispersion were included, one would be able to give valuable
information for the plasmon wave guiding. Knowing that the framework
of this theory allows to include it \cite{Mayergoyz05}, it will be a
further point of interest in the future.

\appendix
\section{General derivation of a gradient formulation of the laplace integral equation}

Let us consider the partial differential equation
\begin{eqnarray}
L(w)=0  ~in ~B,
\end{eqnarray}
where B is a region in 3D-space with surface S, and L is a linear elliptic differential operator operating on a sufficiently smooth function w
 of 3D-variables ${\bf x}$. The Green's reciprocal identity for the operator L can be written:
\begin{eqnarray}
\int_{B} [u L(v) - v \bar{L}(u) ] dB = \int_S
\left(u\frac{\partial v}{\partial n} -v\frac{\partial u }{\partial
n}\right) dS \label{eqgreenidentity},
\end{eqnarray}
in which u and v are arbitrary functions, $\bar{L}$ is the operator adjoint to $L$ and $n$ is the outward normal to S. Let us choose v as the
fundamental solution $G({\bf x},{\bf x'})=1/4\pi |{\bf x}-{\bf x'}| $ (in this case $L=\nabla^2$) such that:
\begin{eqnarray}
L(G)=-\delta ({\bf x}-{\bf x'})
\end{eqnarray}
where $\delta$   is the Dirac delta function. The integrands in Eq.(A2) can be expanded as an appropriate limit process \cite{Muskhelishvili53}. We therefore consider a point ${\bf x}={\bf x}_o$ on the boundary S and let there be a sphere around this point separating a volume $B_\rho$  and surface $S_\rho$  from the rest of B. Let $B_0$ denote what is left of region B with surface $S_0$ comprised of $S_1$ and $S_{\rho}$  according to the Fig. 4
\begin{figure}[htp]
\centerline{\resizebox{8.5cm}{!}{\includegraphics{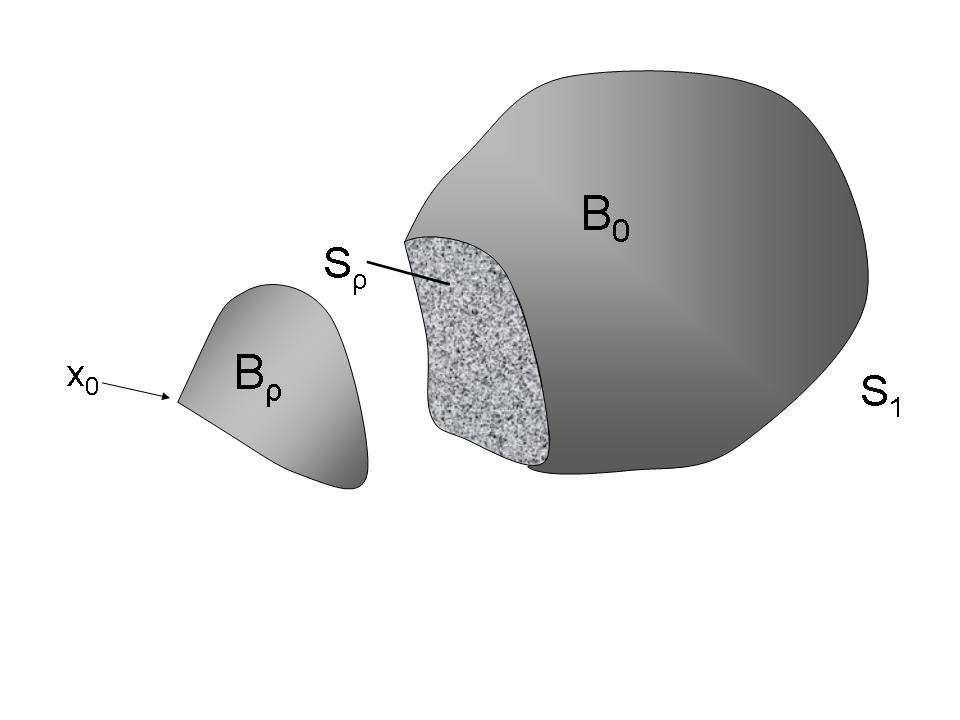}}}
\caption{Segment $B_{\rho}$ and the remainder $B_0$ of the region $B$.}\label{hartmann}
\end{figure}

Evaluating Eq.(A3) for $B_0$ and $S_0$ , which does not contain the point ${\bf x}_0$, only the surface integral is left:
\begin{eqnarray}
0=\int_{S_0} \left(G\frac{\partial u}{\partial n} -u\frac{\partial G }{\partial
n}\right) dS(x')
\end{eqnarray}
The first term of Eq. (A4) has a singularity which is integrable, while the second term has to be treated with more care. Splitting the integral over $S_0$ as $S_1$
and $S_{\rho}$  and adding/subtracting $u({\bf x}_0)$, and in the limit $\rho \rightarrow 0$, we get the main result in \cite{HARTMANN81}(for ${\bf x}_0$ on the boundary):
\begin{eqnarray}
c({\bf x}_0)u({\bf x}_0)=\int_S \left(G\frac{\partial u}{\partial n} -u\frac{\partial G }{\partial
n}\right) dS(x')
\end{eqnarray}
where the second integral is to be interpreted in Cauchy Principal Value sense. Obviously by setting $u=1$ we have:
\begin{eqnarray}
c({\bf x}_0)=\int_S \left( -\frac{\partial G }{\partial
n}\right) dS(x')=\frac{\Omega({\bf x}_0)}{4\pi}
\end{eqnarray}
where we have used  that $\Omega({\bf x}_0 )$   is the solid angle pertaining to the point  ${\bf x}_0$.

\begin{acknowledgments}
The work of R.V. has been supported by the EU Project {\it Nanomagna} under contract NMP3-SL-2008-214107. The work of JIJ has been supported by the Basque Departamento de Educaci\'on, Universidades e Investigaci\'on, the University of the Basque Country UPV/EHU (Grant No. IT-366-07) and the Spanish Ministerio de Ciencia e Innovaci\'on (Grant No. FIS2010-19609-C02-02). PA acknowledges support from Swedish Foundation for Strategic Research - project metamaterials. R.V. wants to thanks specially A. Rivacoba, J. Aizpurua, and P.M. Echenique for useful discussions.
\end{acknowledgments}

\end{document}